\documentclass{acm_proc_article-sp}

\iffalse
\ifCLASSINFOpdf
\else
\fi

\usepackage{ucs}
\usepackage{array}
\usepackage[utf8x]{inputenc}
\usepackage{amsmath}
\usepackage[]{algorithm2e}
\usepackage{algorithmicx}
\usepackage{makecell}
\usepackage{graphicx}
\usepackage{subcaption}
\usepackage{multirow}
\usepackage{layout}
\usepackage{cite}
\usepackage{setspace}
\usepackage{listings}

\usepackage{float}

\floatstyle{ruled}
\newfloat{program}{thp}{lop}
\floatname{program}{Program}

\usepackage{color, colortbl}
\graphicspath{ {diagrams/} }

\definecolor{Gray}{gray}{0.9}
\definecolor{LightCyan}{rgb}{0.88,1,1}

\definecolor{dkgreen}{rgb}{0,0.6,0}
\definecolor{gray}{rgb}{0.5,0.5,0.5}
\definecolor{mauve}{rgb}{0.58,0,0.82}

\usepackage{color, colortbl}

\definecolor{Gray}{gray}{0.9}
\definecolor{LightCyan}{rgb}{0.88,1,1}
\usepackage[T1]{fontenc}
\usepackage{array}
\usepackage{makecell}
\newcolumntype{x}[1]{>{\centering\arraybackslash}p{#1}}

\usepackage{tikz}
\newcommand\diag[4]{%
	\multicolumn{1}{p{#2}|}{\hskip-\tabcolsep
		$\vcenter{\begin{tikzpicture}[baseline=0,anchor=south west,inner sep=#1]
			\path[use as bounding box] (0,0) rectangle (#2+2\tabcolsep,\baselineskip);
			\node[minimum width={#2+2\tabcolsep},minimum height=\baselineskip+\extrarowheight] (box) {};
			\draw (box.north west) -- (box.south east);
			\node[anchor=south west] at (box.south west) {#3};
			\node[anchor=north east] at (box.north east) {#4};
			\end{tikzpicture}}$\hskip-\tabcolsep}}

\usepackage[T1]{fontenc}
\usepackage{mathptmx}

\begin{document}
%
\title{Fine-Grained Energy Modeling for the Source Code of a Mobile Application}

\author{
	\alignauthor Xueliang Li \qquad John P. Gallagher\\
	\email{\{xueliang, jpg\}@ruc.dk} \\
	%
	\affaddr{Roskilde University}	\\
	\affaddr{Denmark}
}


%


\maketitle

\begin{abstract}
Energy efficiency has a significant influence on user experience of battery-driven devices such as smartphones and tablets. The goal of an energy model for source code is to lay a foundation for the application of energy-saving techniques during software development. The challenge is to relate hardware energy consumption to high-level application code, considering the complex run-time context and software stack. Traditional techniques build the energy model by mapping a hardware energy model onto software constructs; this approach faces obstacles when the software stack consists of a number of abstract layers. Another approach that has been followed is to utilize hardware or operating system features to estimate software energy information at a coarse level of granularity such as blocks, methods or even applications. In this paper, we explain how to construct a fine-grained energy model for the source code, which is based on "energy operations" identified directly from the source code and able to provide more valuable information for code optimization. We apply the approach to a class of applications based on a game-engine, and explain the wider applicability of the method.




\end{abstract}


%

\section{Introduction}

In February 2015, the penetration of smartphones was about 75\% in the U.S. This figure is still growing.
 With the improvement of hardware processing capability and software development environment, applications are becoming much heavier and more PC-like. At the same time, users are frustrated by limited battery capacity --applications running in parallel could easily drain a fully-charged battery within 24 hours.

Furthermore,  current software development is performed in an energy-oblivious manner. Throughout the engineering life-cycle, most developers and designers are blind to the energy usage of their code. On the other hand, it has been estimated that energy saving by a factor of as much as three to five could be achieved solely by software optimization \cite{Edwards:lssmlps}. 
To realize this, the first step is to understand the energy attributes of source code at different levels of granularity and from different points of view.

Energy modeling of software needs to bridge the gap between high-level source code and low-level hardware, where energy is consumed, in order to enable the energy accounting of code. However, traditional bottom-to-top modeling techniques \cite{Tiwari:power_analysis_embedded,bran:instruction-level_model,gangqu:function-level_powermodel,Simunic:2000:source_code_optimization} face obstacles when the  software stack of the system consists of a number of abstract layers. On the Android platform, say, the source code is in Java and then translated to Java byte-code, further to Dalvik \cite{Android:Dalvik} byte-code, native code and machine code and finally has chance to execute on the processors and consume energy. Consequently, the modeling task has to characterize the links between all the layers. 

Instead of building a software energy model layer by layer,  
 another approach to acquiring software-level energy information is to use the hardware readings, like CPU state residency, CPU utilization, L1/L2 Cache misses and battery trace, as predictors of software energy use \cite{Dong_selfconstructivemodel,Pathak_whereisenergy,Zhang_onlinepowerestimation,Wang_batterytrace}. However, they are only capable of obtaining energy information at a coarse level of granularity such as blocks, methods or applications. Two pieces of work \cite{HaoShuai:2013:EstMobileApp,sourceline_energy} result in source-line energy information. The former requires low-level energy profiles. The latter employs accurate measurement to attain the energy consumption of source lines.

The main contribution of this paper is to develop a fine-grained software energy model based on "energy operations", which are more fine-grained than source lines. Rather than building the software energy model from bottom to top, we identify the basic energy operations (such as multiplications, comparisons and method invocations) directly from source code and find their correlations to the energy cost by analyzing a diversity of well-designed execution cases. The resulting energy model implicitly includes the effect of all the layers of the software stack down to the hardware.
  
The key contributions of this work are the followings:
\begin{itemize}
	\item A source-code energy model based on fine-grained energy operations, from which energy operations of source lines, blocks, and etc. can be derived.
	\item An approach in which an explicit low-level energy model or a hardware profile is not required, since comprehensive information is identified straight from the source code and statistical analysis of a wide range of execution cases. 
\end{itemize}

The advantages of the operation-based source-level model are listed below:

\begin{itemize}
	\item In comparison with the model for Java or Dalvik instructions, the source-level model provides source-code-related information that is easier to interpret by the developer who plays a significant role in code refactoring.
	\item The operation-based model is able to generate operation-level information, which tells valuable clues on how to make the code more energy-efficient, as shown in Section \ref{Section:code_optimization_by_accounting}.
\end{itemize}
Our target platform is an Android development board with two ARM quad-core CPUs, and the source code in our study is a game engine used in games, demos and other interactive applications. 
The result shows that the model's inference accuracy achieves about 85.0\%. Even though the error looks not tiny, the model is able to produce  more comprehensive energy information for code optimization than coarse-grained models or techniques could provide. 


The generality of the approach goes beyond the boundaries of the case study described here.
Firstly, the energy model constructed can be used in developing
the large class of applications which are based on the game-engine, comprising many interactive applications with rich user interfaces.  Secondly, the modeling methodology is applicable to all kinds of applications. The choice of energy operations is dependent only on the Java source language, and the techniques for designing test cases and regression analysis can be applied to other application domains.


In Section \ref{Section_basicEnergyOp}, we introduce the approach of identifying energy operations from the source code. The architectural setup and the design of execution cases are detailed in Section \ref{Section_experimentsetup}. We elaborate upon the data collection and the model construction separately in Section \ref{Section_dataCollection} and Section \ref{Section:Model}, based on which the fine-grained energy accounting is shown in Section  \ref{Section_Analysis}.

\section{Basic Energy Operations}\label{Section_basicEnergyOp}

There are two reasons why we choose to build the source code energy model based on "energy operations". Firstly, an energy operation is "atomic", by which we mean that all the statements, source lines, blocks and methods are made up of a certain number of kinds of operations (in the experiment, we have 120 operations). Secondly, it is fine-grained. Energy information at the level of source lines or methods is useful; however, information at source line level could not distinguish energy consumption of two operations in the same source line, for example. 



   
   \begin{table}
   	\centering
   	\caption{Examples of Energy Operations\label{EnOps}}
   	\begin{tabular}{ll} 
   		\hline
   		Operation &  \textit{Identified where:} \\ \hline
   		{Method Invocation} & \textit{ one method is called}\\ 
   		
   		Parameter\_Object &  \textit{ Object is one parameter of the method}\\ 
   		
   		Return\_Object &  \textit{ the method returns an Object}\\ 
   		
   		Addition\_int\_int & \textit{ addition's operands are integers }\\ 

   		Multi\_float\_float & \textit{ multiplication's operands are floats }\\
   		Increment & \textit{ symbol "$++$" appears in code}\\ 
   		And & \textit{ symbol "$\&\&$" appears in code}\\ 
   		Less\_int\_float & \textit{ "<"'s operands are integer and float}\\
   		
   		Equal\_Object\_null & \textit{ "=="'s operands are Object and null}\\ 
   		
   		Declaration\_int & \textit{ one integer is declared}\\
   		
   		Assign\_Object\_null & \textit{ assignment's operands are Object and null}\\
   		Assign\_char[]\_char[] & \textit{ assignment's operands are arrays of chars}\\
   		
   		Array Reference &  \textit{ one array element is referred}\\
   		Block Goto & \textit{ the code execution goes to a new block}\\

   		\hline
   	\end{tabular}
   \end{table}



Energy operations are identified directly from source code. The enumeration of the operations is inspired by Java semantics \cite{Bogdanas_Semantics}, which specifies the operational meaning, or behavior, of the Java language, which is the target language in the experiment.
We intuitively identify semantic operations that perform operations on the state and may be energy-consuming, and let them be our energy operations. 
Ones that have little or no energy effect will automatically be identified by the regression analysis in the later stage of the analysis.
Table \ref{EnOps} lists 14 representative operations out of a total of 120 in the experiment.
They include arithmetic calculations like \textit{Multi\_float\_float}, \textit{Addition\_int\_int}, in which operands types are explicit, as well as \textit{Increment} whose operand is implicitly an integer.   Boolean operations and comparisons, such as \textit{And}, \textit{Less\_int\_float} and \textit{Equal\_Object\_null} also form one major part. \textit{Method Invocation} and \textit{Block Goto} are important for the control flow which plays a key role in the execution of the code. Assignments and \textit{Array Reference} will unexpectedly take a significant amount of the application's energy consumption, as will be shown in Section \ref{Section_Analysis}. 
\begin{table}
	\centering
	\caption{Examples of Library Functions\label{Libaray_functions}}
	\begin{tabular}{ll} \hline
		Class & $\qquad$Function \\ 
		\hline
		ArrayList & \textit{add, get, size, isEmpty, remove}   \\ 
		
		& \textit{glBindTexture, glDisableClientState } \\
		& \textit{glDrawElements, glEnableClientState} \\
		GL10     & \textit{glMultMatrixf, glTexCoordPointer} \\
		& \textit{glPopMatrix, glPushMatrix} \\
		& \textit{glTexParameterx, glVertexPointer}\\
		Math    & max, pow, sqrt, random  \\
		
		FloatBuffer  & \textit{position, put} \\
		\hline
		\end{tabular}
\end{table}

The application also employs a diversity of library functions that may be written in different languages and at lower levels of the software stack. On the other hand,  usually a limited number (67 in the experiment) of library functions are frequently called in one application. So we treat them as basic modeling units. The examples of highly-used library functions in the experiment are shown in Table \ref{Libaray_functions}. For instance, the functions in the class of \textit{GL10} are responsible for graphic computing.

\section{Experimental Setup}\label{Section_experimentsetup}

In this section, we will introduce the setup of the target device and the employed source code. We also explain the design principles of the execution cases.

\subsection{Target Device}\label{Section_target_measurement}
Experimental target: we employ an Odroid-XU+E development board \cite{target:odroid} as the target device. It possesses two ARM quad-core CPUs, which are Cortex-A15 with 2.0 GHz clock rate and Cortex-A7 with 1.5 GHz. The eight cores are logically grouped into four pairs. Each pair consists of one big and one small core. So from the operating system's point of view there are four logic cores. In our experiment, we turn off the small cores and run workload on big cores at a fixed clock frequency of 1.1 GHz. We do this in order to remove the influence of voltage, clock rate and CPU performance on the power usage.   

Power Reading Script: Odroid-XU+E has built-in sensors to measure the voltage and current of CPUs with a frequency of 30 Hz and updates the samples in a log file. We wrote a script to obtain the samples from the file. During execution we run the script on an idle core to minimize its influence on the application.   

Note that the power monitor gives two sequences of power samples: one is for the big cores and the other is for the small cores. We pick the sequence of power samples of the big cores, because we only run workload on them.


\subsection{Target Source Code}

The target source code is the Cocos2d-Android \cite{code:cocos2d} game engine, a framework for building games, demos and other interactive applications. It also implements a fully-featured physics engine. Games are increasingly popular on mobile phones and include more and more fancy and energy-consuming features, requiring high CPU performance. This paper demonstrates the energy modeling for the source code of this game engine, but the methodology to be seen later is applicable for all kinds of applications in principle.


\subsection{Design of Execution Cases }\label{Section_sourcecode_casedesign}

The execution cases whose energy usage is measured and analyzed represent typical sequences of actions during  game, including user inputs. We focus on a \texttt{Click \& Move} scenario, in which the sprite (the character in the game) moves to the position where the tap occurs. To simulate the game scenarios under different sequences of user inputs, we script with the Android Debug Bridge \cite{adb:android} (ADB) , a command line tool connecting the target device to the host, to automatically feed the input sequences to the target device.
   
In order to obtain a more varied set of execution cases, we vary the executions of individual basic blocks in the code. This is achieved by systematically removing a set of blocks for each execution case, using the control flow graph obtained using the Soot tool \cite{soot:callgraph}. We ensure that each block could be removed in some execution case.
Thus an execution case is made up of one user input sequence and one set of basic blocks.

\section{Data Collection}\label{Section_dataCollection}

In this section, we describe the collection of data on the number of times each operation executes and the energy consumption of an execution case, based on which we construct the energy model. 

\begin{figure}
	\centering
	\begin{subfigure}[b]{0.24\textwidth}
		\centering            
		\includegraphics[width=\textwidth]{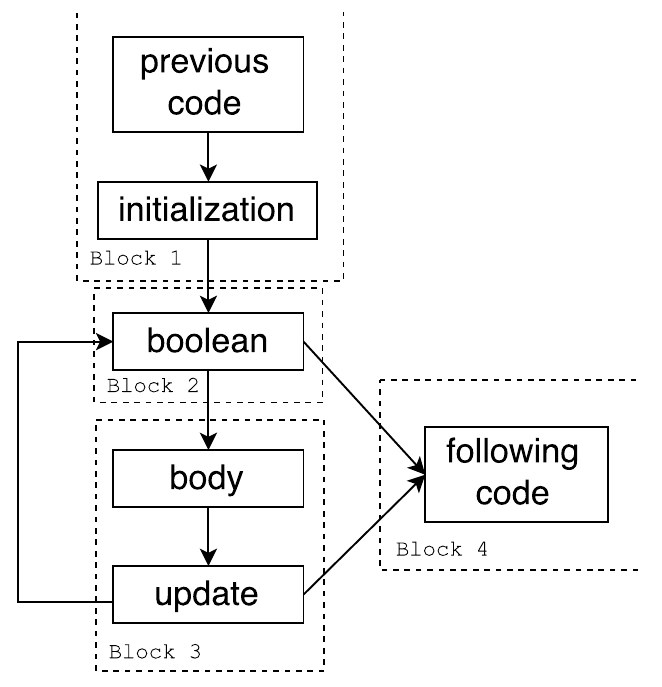}
		\label{fig:for_loop}
	\end{subfigure}%
	\begin{subfigure}[b]{0.24\textwidth}		
		\centering
		\raisebox{5mm}{\includegraphics[width=\textwidth]{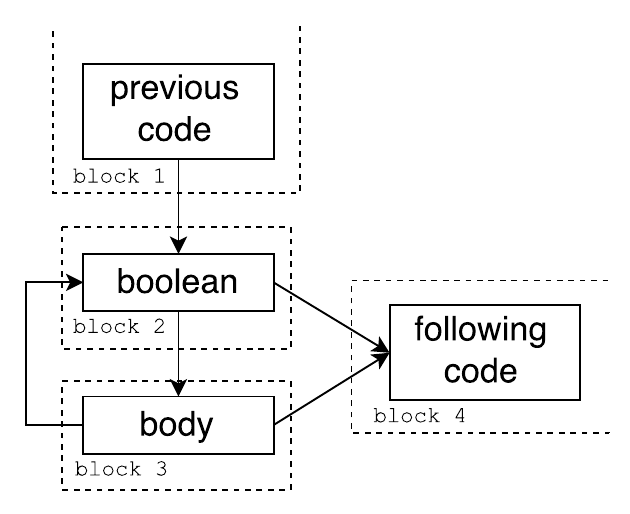}}
		\label{fig:while_loop}
	\end{subfigure}
	\caption{Block division of for and while loops in control flow graph.}\label{fig:block_division}
\end{figure}

\subsection{Number of Executions of Operations}
To obtain the number of times that each operation executes in an execution case, we need to determine at which level of granularity to track the execution. We choose the level of "blocks". A block is a sequence of consecutive statements, without loops or branches.      
It is sufficient to track block executions, since if one part of a block is processed, the rest certainly will be processed as well.   

We could consider collecting data at other levels of granularity. Tracing individual statements might overload the capacity of the target device. On the other hand, methods or classes are unsuitable execution units, since we cannot determine which parts of the method or class will be active during the execution, and this information about energy operations is lost.

We then divide the source code into blocks.
For individual syntactic structures, we deal with block division case by case. \texttt{For} loops and \texttt{while} loops are handled as shown in Figure \ref{fig:block_division}. In a \texttt{for} loop, the header usually has three segments which are \textit{initialization}, \textit{boolean} and \textit{update}. They are divided into three different blocks. Similarly, we set the \texttt{while} header itself as a block ("block 2" in Figure \ref{fig:while_loop}). In order to build the log, we instrument the source code with a log instruction at the beginning of each block.

\begin{figure}
	\centering
	\includegraphics[width = 0.35\textwidth]{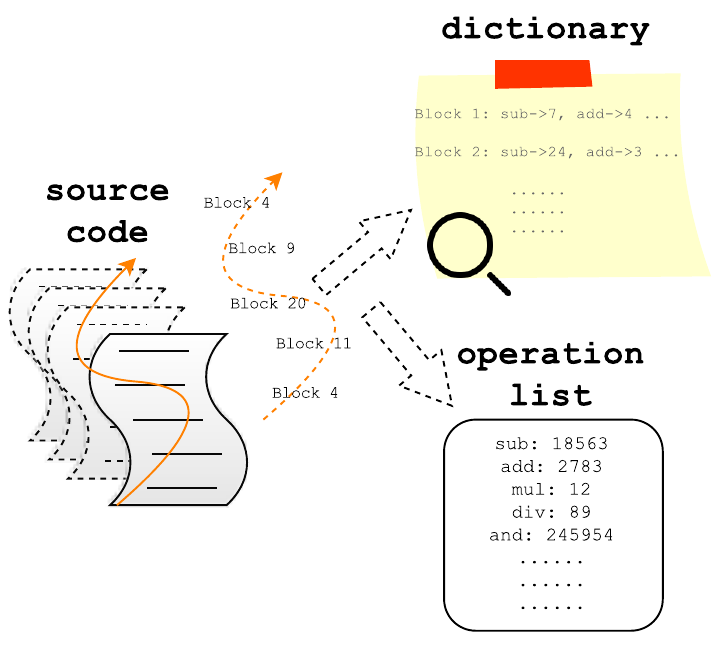}
	\caption{The flow of the operation-execution data collection.}\label{fig:data_collection_flow}
\end{figure}

The generic view of the collection of  the operation-execution data is displayed in Figure \ref{fig:data_collection_flow}. 
We build a dictionary showing, for each block, the number of occurrences within it of each energy operation, such as those in Table \ref{EnOps}.  This dictionary is built using a parser that traverses all the blocks in the code. 

Then, using the log file recording the processed blocks, together with the dictionary,  we can sum up the number of times that each energy operation is executed during an execution case. To be more precise, let $B_i$ be the number of times that the $i^{th}$ block is executed (this is obtained from the log file).  Let $O_{i,j}$ be the number of occurrences of operation $j$ in block $i$ (this is obtained from the dictionary).  Then the total number of executions of the $j^{th}$ operation is $\sum_{i=1}^{n} (B_i * O_{i,j})$, where $n$ is the total number of blocks.

\subsection{Energy Approximation from Power Samples}

We write a script to obtain the power samples from the built-in measurement component with a frequency of 30 Hz. The power samples are the discrete values sampled from the power trace; we approximate energy consumption by calculating Equation (\ref{Energy_Equation}): $p=\emph{power}(t)$ is the power trace, that is, the continuous power-vs-time function; $\emph{power}(t_i)$ is the power sample at time-stamp $t_i$; $\Delta_i$ equals to $t_i - t_{i-1}$, which is the interval between two sequential samples.

\begin{equation}\label{Energy_Equation}E=\int_{t_0}^{t_n}\textit{power}(t)\,dt
 \;\approx\sum_{i=1}^{n}\textit{power}(t_i)\cdot\Delta_i\; \end{equation}
 
\begin{displaymath}\quad where
\quad t_0\le t_1\le t_2 \cdots \le t_{n-1}\le t_{n}  \end{displaymath}

\subsection{Challenges in Practice }

\textbf{Measurement limitation:} the sampling rate of the built-in power monitor is 30 Hz. However, the instruction execution rate is about several million per second. That means, one power sample measures the energy cost of hundreds of thousand instructions. Even  though the state of the art of the power measurement can reach a sampling rate of tens of KHz \cite{Jiang_bestpowermeter}, one power sample still includes up to thousands of instructions. 

To deal with this problem, we first lengthen the sessions of all the execution cases to above 100 seconds, and then run each case for ten times to calculate their average energy cost. Compared with the execution cases that only run once with sessions around one second, this approach can reduce the error of measuring energy consumption of the code by three orders of magnitude.  
\\\\
\textbf{Run-time context:} during the running of the application, the Dalvik virtual machine performs garbage collection, which is not part of the application and still could be included in the power samples. 

The Dalvik virtual machine produce time-stamp logs when launching the garbage collection procedure. We consider the garbage collection as one library function, so it will be integrated in the model.
\\\\
\textbf{Code instrumentation and power reading script:} although the instrumentation is at block level rather than statement level, its impact on energy consumption is still not negligible and its cost is as much as 50\% of the application's energy consumption itself.  Also, the energy cost of the power reading script is up to 5\% of the application's consumption.

We followed three experimental principles to address this problem. Firstly, for each execution case, the log of the execution path and of the power samples are separated into two separate runs. In the first round, we record the execution path without reading power samples. In the second round, we only trace power and disable the instrumented log instructions. So for each execution case, the instrumentation for logging the execution path will not influence the power samples.

Secondly, in each of the two runs, the main process of  the application is allocated to one CPU core, while the thread logging execution path or power samples is allocated to another CPU core, minimizing effects due to interaction of the threads. 

Thirdly, we design one "idle execution case" paired with each execution case;  this only runs the power reading script without the application. By this means we can get the energy consumption of the main application process by excluding the cost of the "idle execution case" from the execution case. Note that the durations of execution cases are different, so we need to have a distinct "idle execution case" for each execution case.

In summary, each execution case will be run 21 times: once for tracing the execution path; ten times for calculating the average energy consumption of the "idle execution case", and ten times for calculating average energy consumption of the execution case.

\section{Model Construction}\label{Section:Model}


 The entire energy consumption is composed of three parts: the cost of energy operations, the cost of library functions and the idle cost. The aimed model is formalized in Equation (\ref{Energy_Model}). The cost of energy operations is the sum of $  Cost_{op_i} \cdot  N_e(op_i) $ (the cost of one operation multiplied by the number of its executions), where $op_i \in Energy\,Ops$. $Energy\,Ops$ is the set containing all the operations. The cost of library functions is the sum of $Cost_{func_i} \cdot  N_e(func_i)$ (the cost of one library function multiplied by the number of its executions), where $func_i \in Lib\,Funcs$. $Lib\,Funcs$ is the set of library functions.
The $Idle\; Cost$ is the energy consumption of the "idle execution case". The lengths of case sessions are varying, so the $Idle\; Cost$ is different for each execution case.

\begin{equation}\label{Energy_Model}
E = \sum_{}^{op_i \in Energy\,Ops} Cost_{op_i} \cdot  N_e(op_i) \qquad\qquad \end{equation}
\begin{displaymath}
+ \sum_{}^{func_i \in Lib\,Funcs} Cost_{func_i} \cdot  N_e(func_i) + Idle\;Cost
\end{displaymath}


The model construction is based on regression analysis, finding out the correlation between energy operations and their costs from the data obtained in the execution cases. We set out the collected data in the matrices in Equation (\ref{equation_matrices}). The leftmost matrix ($N$) contains the execution numbers of $l$ operations (including energy operations and library functions) in $m$ execution cases, acquired as shown in Section \ref{Section_dataCollection}. Each row indicates one execution case. Each column represents one operation. The vector ($\vec{cost}$) in the middle contains the costs of $l$ operations, which are the values we are aiming to estimate. The vector ($\vec{e}$) on the right of the equal mark contains the measured entire energy costs of the execution cases. So for each execution case, the entire energy cost is the sum of the costs of operations. It should be noticed that the energy costs $\vec{e}$ exclude the $Idle\; Cost$ which is measured when no application workload is being processed. \\

\begin{equation}\label{equation_matrices}
\begin{pmatrix} 
n_1^{(1)}  & n_2^{(1)} & ... & n_l^{(1)}\\ 
n_1^{(2)}  & n_2^{(2)}  & ... & n_l^{(2)}\\
&        ...& ...    &          \\
n_1^{(m-1)}  & n_2^{(m-1)}  &  ... & n_l^{(m-1)}\\
n_1^{(m)}  & n_2^{(m)}  &  ... & n_l^{(m)}
\end{pmatrix} \times
\begin{pmatrix}
cost_1\\
cost_2\\
...\\
cost_l
\end{pmatrix} =
\begin{pmatrix}
e_1\\
e_2\\
...\\
e_{m-1}\\
e_m
\end{pmatrix}\\
\end{equation}

Inevitably, the power samples are not absolutely accurate. Furthermore, the energy model in reality is unlikely to be completely linear. For these reasons Equation (\ref{equation_matrices}) may be unsolvable, that is, the vector $\vec{e}$ is out of the column space of $N$. We thus employ the gradient descent algorithm \cite{Ng:Machinelearning} to compute the approximate values of $\vec{cost}$.

The elements of $\vec{cost}$ are randomly initialized and then improved by the gradient descent algorithm iteratively. We first introduce the error function $J$ (computed by Equation (\ref{equation_ErrorFunction})) which indicates the quality of the model. The smaller $J$ is, the better the model is. $\vec{n^{(i)}}$ is the $i^{th}$ row in $N$, $\vec{cost}$ is the middle vector above. $\vec{n^{(i)}}\times\vec{cost}$ is the estimated energy cost for the $i^{th}$execution case, $e^{(i)}$ is its observed energy cost. $J$ first computes the sum of the squared values of the estimate errors of all the execution cases, which is afterwards divided by $2m$ to get the average value. 

\begin{equation}\label{equation_ErrorFunction}
J(cost_1, cost_2,...cost_l)=\frac{1}{2m}\sum_{i=1}^{m}(\vec{n^{(i)}}\times\vec{cost}-e^{(i)})^2
\end{equation}

\begin{equation}\label{equation_updatecost}
cost_j := cost_j - \alpha\frac{\partial J(cost_1,...cost_j,...cost_l)}{\partial cost_j}
\end{equation}
\begin{displaymath}
= cost_j - \alpha \frac{1}{m} \sum_{i=1}^{m}(\vec{n^{(i)}}\times\vec{cost}) \cdot n_j^{(i)}
\end{displaymath}
\begin{displaymath}
j = 1,2,...l
\end{displaymath}

The idea of gradient descent is to minimize $J$ by repeatedly updating all the elements in $\vec{cost}$ with Equation (\ref{equation_updatecost}) until convergence. The partial derivative of the function $J$ on $cost_j$ gives the direction in which increasing or decreasing $cost_j$ will reduce $J$. Every element ($cost_j$) of $\vec{cost}$ is updated one by one in each iteration. The value $\alpha$ determines how large the step of each iteration is. If it is too large, the extremum value will possibly be missed; if too small, the minimizing process will be rather time-consuming. It needs to be manually tuned. Theoretically, the gradient descent algorithm could only find the local optima. In practice, we randomly set the values in $\vec{cost}$ and restart the entire gradient decent procedure for several times to look for the global optima.


\begin{table}
	\centering
	\caption{Correlation Coefficient in Cross Validation\label{Cross Validation_correlation}}
	\begin{tabular}{c|c|c|c|c} \hline
		\diag{.1em}{2.13cm}{$\quad$Set}{$\;$Round}& 1st & 2nd & 3rd & 4th \\ 
		\hline
		Training set   & 0.84 & 0.81 & 0.84 & 0.84  \\
		\hline
		Validation set & 0.91 & 0.91 & 0.88 & 0.88   \\
		\hline\end{tabular}
\end{table}

\begin{table}
	\centering
	\caption{NMAE in Cross Validation\label{Cross Validation}}
	\begin{tabular}{c|c|c|c|c} \hline
		\diag{.1em}{2.13cm}{$\quad$Set}{$\;$Round}& 1st & 2nd & 3rd & 4th \\ 
		\hline
		Training set   & 15.7\% & 14.1\% & 16.3\% & 15.2\%  \\
		\hline
		Validation set & 9.3\% & 15.7\% & 9.5\% & 11.6\%   \\
		\hline\end{tabular}
\end{table}

\begin{figure}
	\centering
	\includegraphics[width = 0.45\textwidth]{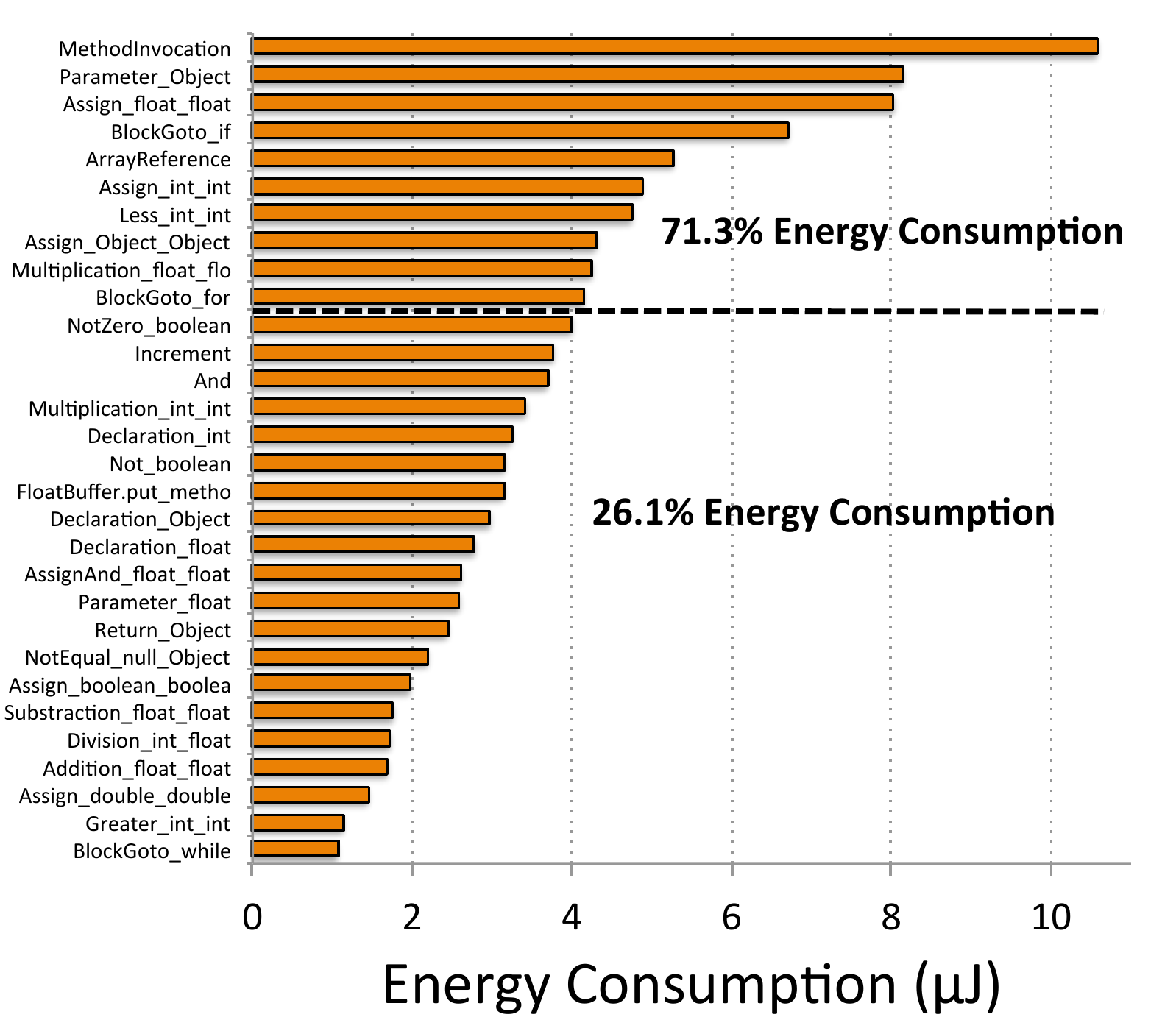}
	\caption{The top 30 energy consuming operations.}\label{fig:op_rank}
\end{figure}

To validate the model, we apply the four-round cross validation procedure: the set of execution cases are randomly divided into four subsets; in each round, one of them is chosen to be the validation set and the others together to be the training set. We utilize two statistical criteria to assess our model. The first one is the correlation coefficient ($r$) that represents the strength and direction of the linear relationship between estimated and measured values. Table \ref{Cross Validation_correlation} presents $r$ in training and validation sets in the four rounds, which shows $r$ is around 0.85 in general, which means the estimated value has a positive and strong relationship with its corresponding measured value.  

The other criterion is the Normalized Mean Absolute Error (NMAE). The NMAE is a well-known statistical criterion that indicates how well the estimated value matches the measured one. It is computed by Equation (\ref{equation_NMAE}), the mean value of normalized difference between the predicted energy cost $\hat{e}$ and the measured cost $e$. The lower the ratio the better the result. In Table \ref{Cross Validation}, we can see NMAE of the model in training and validation sets in the four rounds. The NMAE in training sets ranges from 14.1\% to 16.3\%, and in validation sets from 9.3\% to 15.7\%. The NMAEs are around 15.0\%, which means the model's inference accuracy is around 85.0\%.


\begin{equation}\label{equation_NMAE}
NMAE= \frac{1}{n} \sum_{i=1}^{n} | \frac{\hat{e^{(i)}} - e^{(i)}}{e^{(i)}} |
\end{equation}

\section{Energy Accounting}\label{Section_Analysis}
The energy model for the application source code based on energy operations facilitates comprehensive energy accounting at different levels of granularity and from various viewpoints. In this section, we will see the rank of the most expensive operations, and the contributions of different operations to the energy consumption of each block.  

\begin{figure*}
	\centering
	\begin{subfigure}[b]{0.44\textwidth}
		\centering            
		\includegraphics[width = \textwidth]{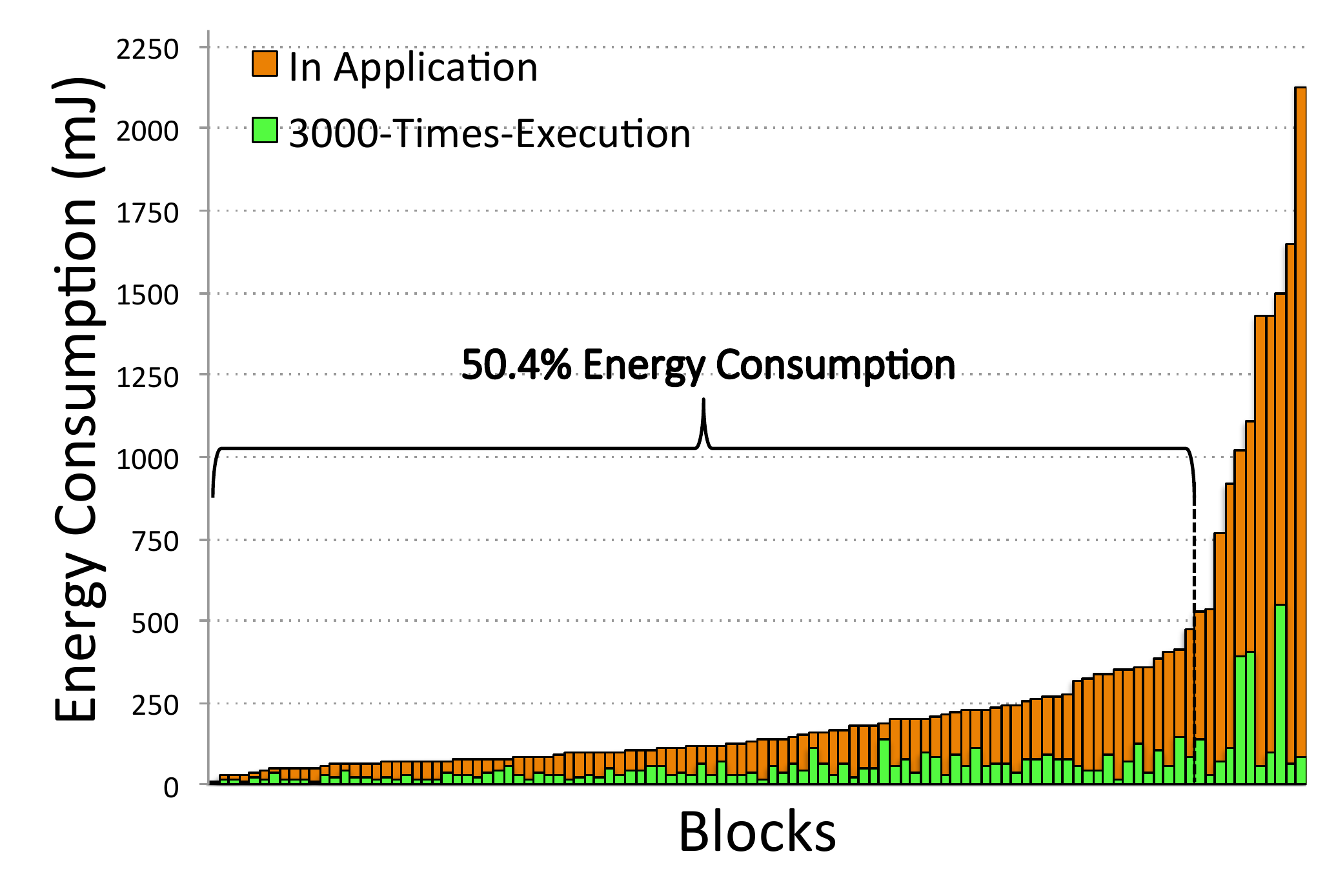}
		\label{fig:blocks_in_program}
	\end{subfigure}
	\begin{subfigure}[b]{0.44\textwidth}
		\centering
		\includegraphics[width=\textwidth]{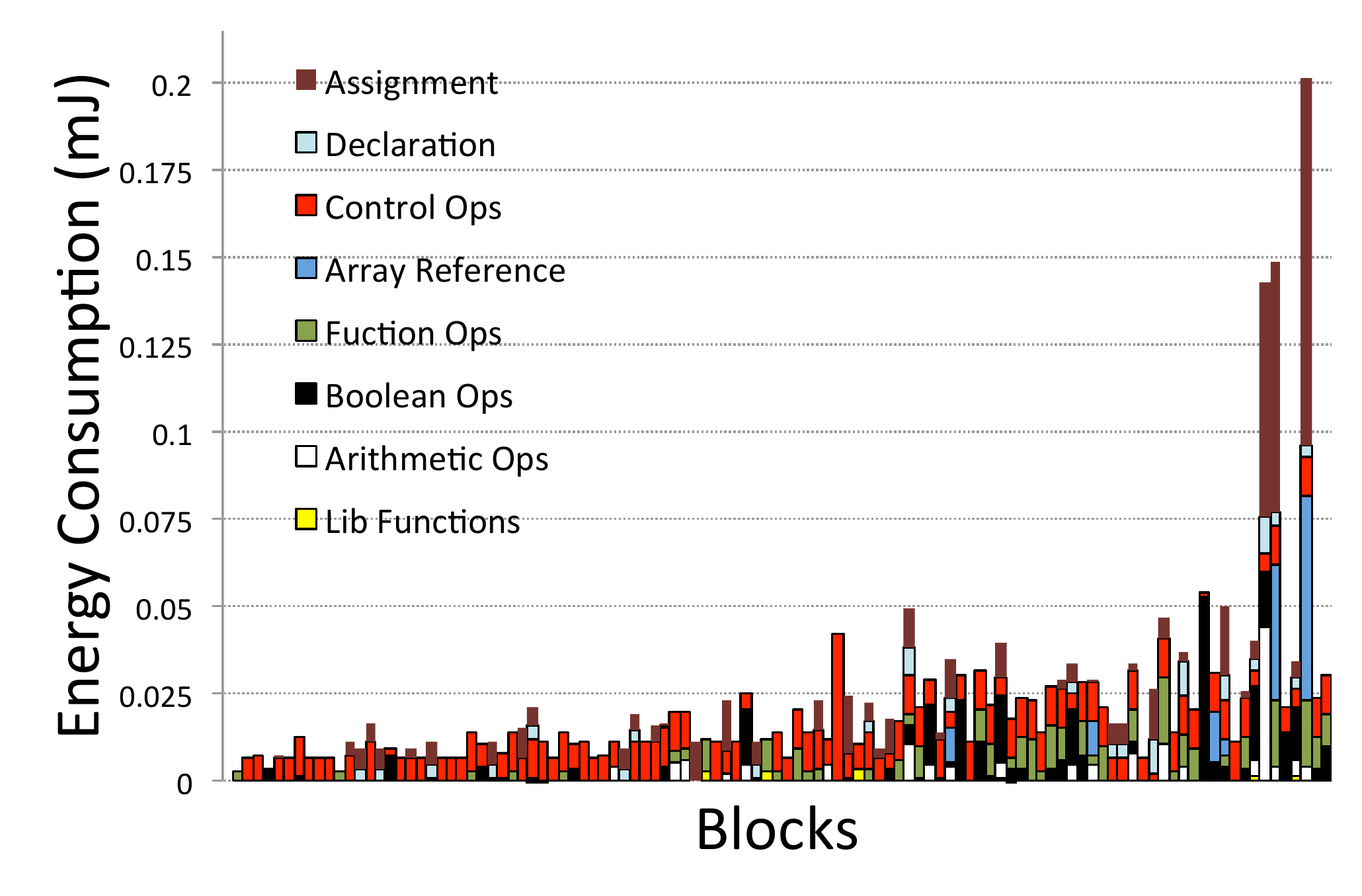}
		\label{fig:op_in_block}
	\end{subfigure}
	\caption{Energy distribution on blocks and operations. Blocks are sorted by the order of their run-time energy costs "In Application".}\label{fig:block}
\end{figure*}

\subsection{Operation Level}

Figure \ref{fig:op_rank} shows the top 30 energy consuming operations, which are ranked by their single-execution energy costs. "71.3\% Energy Consumption" presents the percentage of sum of costs of top 10 operations in the total cost, considering their different numbers of executions in the Click \& Move scenario. "26.1\% Energy Consumption" means the percentage of operations from 11th to 30th. The percentages indicate that the energy-usage of the code is largely determined by a relatively small number of operations. It is because these operations are frequently used and meanwhile expensive
themselves. The 30 operations out of 187 (including library functions) take up 97.4\% of the whole cost of the code, in which the top 10 consumes the major part with a percentage of 71.3\%.

Usually, it is supposed that the sophisticated arithmetic operations, such as multiplications and divisions, should be the most costly. However, the result shows that \textit{Method Invocation} ranks the highest. This is due to a sequence of complex processes to fulfill \textit{Method Invocation}, for example, most of the calls in Java are virtual invocations which are dispatched on the type of the object at run-time and always implicitly passed a "this" reference as their first parameter, no mention of other processes, such as storing the return address and managing the stack frame.
It suggests a trade-off between the structure and the energy saving when writing the code. That means, in certain cases, we could unpack some thin methods that are highly-invoked in the code, at the cost of losing the integrity of the structure of the code to some extent.   

Unexpectedly, only one arithmetic operation, \textit{Multi\_float\_float}, is a member of the top 10. And there are only six arithmetic operations in the top 30. They together cost only 6.1\% of the overall energy consumption of the application, which is contrary to our instincts. 


Later in Section \ref{Section_blocklevelenergy}, we will see that assignments, comparisons and \textit{Array Reference} play significant roles in the overall energy consumption. This is not only because they are frequently used, but also because they are costly as operations themselves, as shown in Figure \ref{fig:op_rank}. 

\textit{Block Goto} operations are expensive as well.
Based on the types of conditionals and loops where "Block Goto" occurs, they are classified into \textit{BlockGoto\_if}, \textit{BlockGoto\_for} and \textit{BlockGoto\_while}. The result shows that they cost different amounts of energy as operations themselves, respectively 6.7 $\mu$J, 4.1 $\mu$J, 1.1 $\mu$J. And together with \textit{Method Invocation}, they take up 37.6\% of the total energy consumption of the application. 

\subsection{Block Level}\label{Section_blocklevelenergy}

In the execution cases, we have 108 active blocks with a wide diversity of energy usage. As shown in Figure \ref{fig:blocks_in_program}, "In Application" means running the \texttt{Click \& Move} scenario with the full set of blocks. The costs of blocks "In Application" are plotted as orange bars. Note that, blocks here obviously have distinct execution times. The cost of a fixed number (3000) of executions of one block are calculated by multiplying its single-execution cost by 3000. This could help us compare the single-execution costs of different blocks. The costs of blocks "3000-Times-Execution" are plotted as green bars.

Similar to energy distribution on operations, only a small number (11 blocks) of all the blocks uses up nearly half of the entire cost, which indicates that putting efforts on optimising a small group of blocks can achieve significant energy-saving.

There are two factors that make one block costly "In Application". The first factor is a large number of executions. For example, the most costly block  "In Application" (the rightmost orange bar in Figure \ref{fig:blocks_in_program}) has a large number of execution times. This block takes only 30.6 $\mu$J for single-execution but 2128.6 mJ when running "In Application". 
The second factor is the energy consumption of the block itself. For example, the three prominent green bars in Figure \ref{fig:blocks_in_program}, whose single-execution costs are 201.5 $\mu$J, 146.9 $\mu$J and 142.8 $\mu$J. We will later zoom in these three blocks to see which operations contribute to their energy costs.


We can further observe the energy proportions of operations in each block in Figure \ref{fig:op_in_block}. To illustrate, operations are grouped into eight classes. Specifically, the "Block Goto" operations and \textit{ Method Invocation} are gathered in \textit{Control Ops}; the parameter passing and the value returns of methods are in \textit{Function Ops}; the comparisons and Booleans are in \textit{Boolean Ops}; all the arithmetic computations are in \textit{Arithmetic Ops}; all the library functions are in \textit{Lib Functions}.

Most of the blocks cost less than 25 $\mu$J for single-execution. In these blocks, \textit{Control Ops} occupy the major part of the energy consumption, in contrast, \textit{Arithmetic Ops} only take a tiny proportion. 

For those three most prominent blocks, assignments and \textit{Array Reference} are the biggest energy consumers. Furthermore one of the three blocks has the largest proportion of \textit{Arithmetic Ops} among all the blocks.  

The most expensive block "In Application" consists of three even parts: \textit{Control Ops}, \textit{Function Ops} and \textit{Boolean Ops}. This block is the main entrance of the game engine to draw and display frames, so its works are conditional judgments and method invocations.    


\subsection{Code Optimization Enabled by Accounting}\label{Section:code_optimization_by_accounting}

The eventual purpose of energy modeling and accounting is to guide the direction of code optimization. And in return, the energy-saving by code optimization is a further validation for the helpfulness of the modeling and accounting. Our latest research presented that, guided by the model, the energy-aware programming approach was adequate to save half of the CPU energy consumption. The general procedures of the approach are as followed:


	\begin{itemize}
		\item We utilize the methodology described in this paper to construct the operation-based source-level energy model, which is achieved by analyzing the data produced in a range of well-designed execution cases .
		\item The model generates energy accounting at operation and block level to capture the key energy characteristics of the code. 
		\item We put efforts on the most costly blocks, where we refactor the code to remove, reduce or replace the expensive operations, meanwhile maintain its logical consistency with the original code. 
	\end{itemize} 


We here take the \texttt{Orbit} scenario as an example, where the character in the game together with the grid background spins in three-dimension space. 


In the \texttt{Orbit} scenario, the block \textit{CCGrid3d.blit().for\_1} dominates the overall energy consumption. 80.9\% of the entire cost is consumed by this block. The second most costly block consumes only 1.3\%. "In Application" here means running the \texttt{Orbit} scenario without removing any block. Later in this section, we only focus on this single block. 




Program \ref{program7} shows the original code of \textit{CCGrid3D.blit().\\for\_1}. In this block, the \textit{Control Ops} (\textit{BlockGoto\_for} and \textit{Field Reference}) use up 35.6\% of the energy; \textit{Boolean Ops} use up 20.5\%; the assignments use up 16.7\%; \textit{Arithmetic Ops} use up 14.0\%; \textit{Lib Functions} use up 13.3\%. We find three easy changes to reduce or replace the pricey operations.\\

\textit{a) Loop-Invariant Code Motion:}
In this block, the value of \textit{vertices.limit()} is the constant 2112; we therefore hoist it outside the loop and replace it with the variable \textit{limit}, as shown in Program \ref{program8}. This change avoids invocations and executions of \textit{vertices.limit()} and at the same time decreases a small amount of \textit{Field Reference}. 

\begin{program}
	\begin{lstlisting}
	for (int i = 0; i < vertices.limit(); i=i+3) {
	mVertexBuffer.put(vertices.get(i)); 
	mVertexBuffer.put(vertices.get(i+1));
	mVertexBuffer.put(vertices.get(i+2));
	}
	\end{lstlisting}
	\caption{The \textbf{original} code of \textit{CCGrid3D.blit().for\_1}}
	\label{program7}
\end{program}

\begin{program}
	\begin{lstlisting}
	int limit = vertices.limit(); //added
	for (int i = 0; i < limit; i=i+24) { //changed
	mVertexBuffer.put(vertices.get(i)); 
	mVertexBuffer.put(vertices.get(i+1));
	mVertexBuffer.put(vertices.get(i+2));
	...                       
	mVertexBuffer.put(vertices.get(i+23));//added
	}
	\end{lstlisting}
	\caption{The changed Program \ref{program7}}
	\label{program8}
\end{program}

\textit{b) Loop Unrolling: }
Also as shown in Program \ref{program8}, we duplicate the loop body eight times, reducing the times of comparisons, \textit{BlockGoto\_for}s, assignments and additions. Note that we set the value of the increment as 24 since 24 is a factor of the \textit{limit}, 2112.  

\textit{c) Full Use of Library Function: }

The job of Program \ref{program7} or Program \ref{program8} is to get all the elements in \textit{vertices} one by one and put them one by one into \textit{mVertexBuffer}.  Program \ref{program7} can be simply replaced by one line: \textit{mVertexBuffer.put(vertices.asReadOnlyBuffer())}. This puts all the elements of \textit{vertices} into \textit{mVertexBuffer}. This change realizes the same functionality using the already existing library function, which is one of the key library functions already compiled into native code. 

\subsection{Evaluation}\label{orbit_evaluation}

\begin{figure}
	\centering
	\includegraphics[width = 0.41\textwidth]{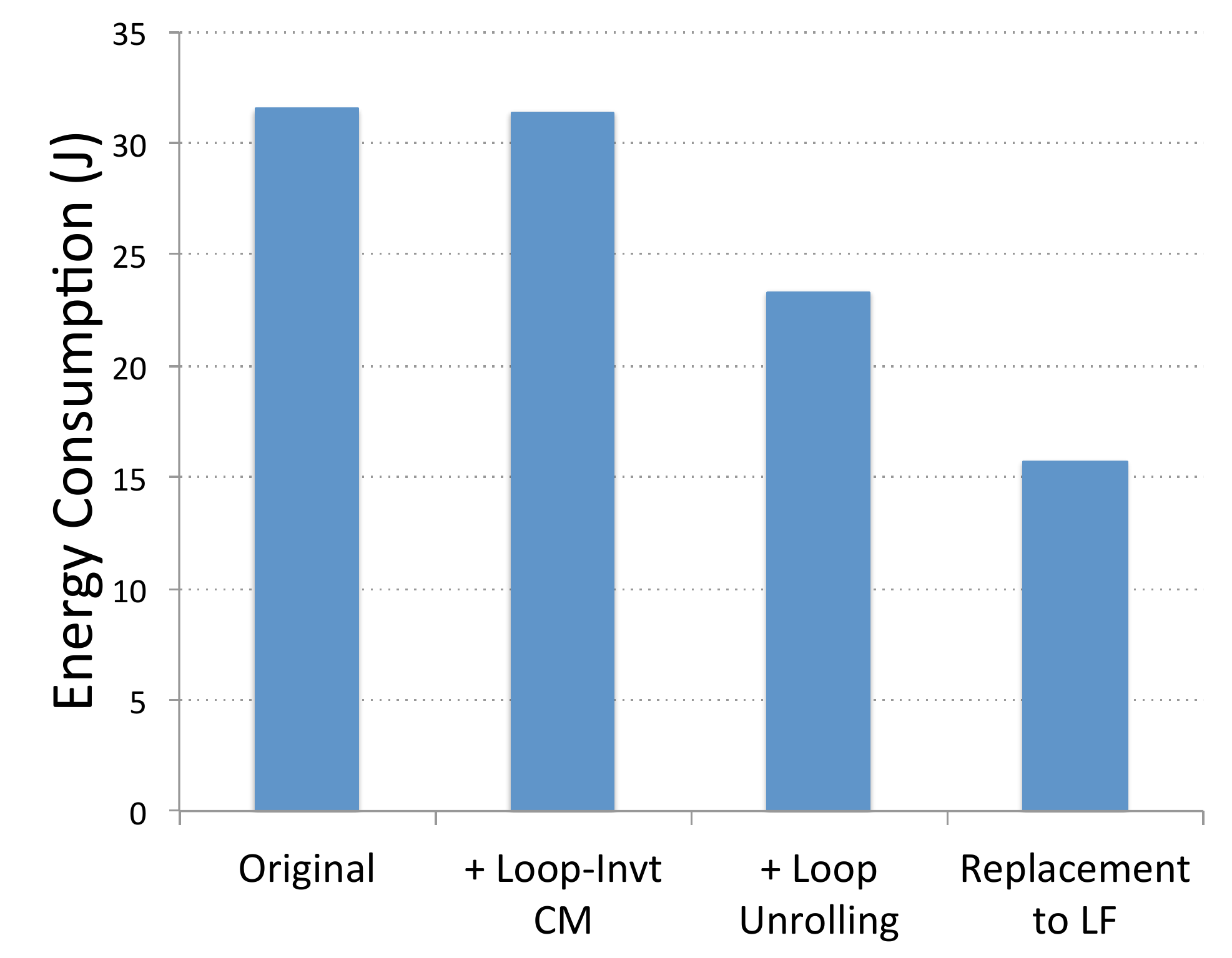}
	\caption{Energy consumption of the code without and with the changes in \texttt{Orbit}.}\label{fig:energy_saving_orbit}
\end{figure}

Figure \ref{fig:energy_saving_orbit} shows the cumulative effects of the code changes on energy consumption. In contrast to the other columns, "\textit{Full-Use LF}" does not take previous changes into account and means only replacing Program \ref{program7} with the built-in library function as stated above. The figure shows that loop-invariant code motion does not gain much energy saving because  \textit{vertices.limit()} is a library function and in addition uses a very small percentage of energy consumption. On the other hand, loop unrolling achieves 25.8\% energy saving due to the reduction of the amount of \textit{Control Ops}, comparisons and assignments, which occupy most of the cost. The most effective change is the replacement to the library function, avoiding the waste of 50.2\% energy use because 
this library function has been compiled into native code before execution, in contrast the Java source code need run-time interpretation which of course incurs an energy cost. The result implies that it is a good idea for developers to make a good use of library functions rather than implementing the same function with Java source code.  The discovery of this source of inefficiency was assisted by the energy accounting.

\section{Related Work}

From the hardware side, initial efforts on energy modeling research have been put on circuits-level (see the survey \cite{Najm_VLSI_level}), gate-level \cite{Najm_gate_level, Marcu_gate_level} and register-transfer-level \cite{cheng_RT_level}. Later,  research focus shifted towards high-level modelings, such as software and behavioral levels \cite{Macii_high_level}.

Energy modeling techniques for software start with the basic instruction level, which calculates the sum of energy consumption of basic instructions and transition overheads \cite{Tiwari:power_analysis_embedded,bran:instruction-level_model}.   
Gang et al. \cite{gangqu:function-level_powermodel} base the model at the function-level while considering the effects of cache misses and pipeline stalls on functions. T. K. Tan et al. \cite{Tan:2001:high-level_softwaremodel} utilize regression analysis for high-level software energy modeling. 

However, the run-time context considered in the above works is unsophisticated, free from user inputs, a virtual machine and so on. Furthermore the software stack below the level that they deal with (such as the level of the basic or assembly instruction) is relatively thin. 

When research is focused  on the energy of mobile applications, the level of granularity of the techniques is increased as well. An important part of such efforts is the use of operating system and hardware features as predictors to estimate the energy consumption at the component, virtual machine and application level  \cite{Dong_selfconstructivemodel, Kansal_powerofvm, Pathak_whereisenergy, Zhang_onlinepowerestimation, Wang_batterytrace, Shye_intowild}.

Shuai et al. \cite{HaoShuai:2013:EstMobileApp} and Ding et al. \cite{sourceline_energy} propose approaches to get source line energy information. The former requires the specific energy profile of the target system, and the workload is fine-tuned. The latter   utilizes advanced measurement techniques to obtain the source line energy cost. 

In contrast to the above approaches, we explore the idea of identifying energy operations and constructing a fine-grained model which is able to capture energy information at a level lower than source line. 


\section{Conclusion}

In this paper, we construct a fine-grained energy model for mobile application source code on the basis of energy operations. We first introduce the energy operations that are identified directly from the source code. The energy operations are employed as the basic units that constitute the overall energy consumption of the source code. We then design a wide diversity of execution cases to generate data about the operation executions and the entire energy consumption. Regression analysis is applied to use the data to estimate the energy consumption of each operation. Finally, we show that the model is capable to capture comprehensive energy features that coarse-grained models or techniques could not shed light on.




\bibliographystyle{abbrv}
\bibliography{DAC2016}  

\end{document}